\documentclass[12pt]{article}

\usepackage{graphicx} 
\usepackage{fullpage}
\usepackage{amsmath}
\usepackage{amssymb}
\usepackage{cite}

\def\eq#1{Eq.~(\ref{#1})}

\newcommand{\secn}[1]{Section~\ref{#1}}

\newcommand{\be}{\begin{equation}}
\newcommand{\ee}{\end{equation}}
\newcommand{\bea}{\begin{eqnarray}}
\newcommand{\eea}{\end{eqnarray}}

\newcommand{\as}{\alpha_s}

\newcommand{\beq}{\begin{eqnarray}}
\newcommand{\eeq}{\end{eqnarray}}
\newcommand{\cK}{\mathcal{K}}

\begin{document}


\titlepage

\phantom{space}

\vspace{1cm}

\begin{center}

{\Large \bf The long road from Regge poles to the LHC\footnote{A contribution 
to the forthcoming volume ``Tullio Regge: an eclectic genius, from quantum gravity 
to computer play'', Eds. L Castellani, A. Ceresole, R. D'Auria and P. Fr\'e, World 
Scientific. }}

\vspace{2cm}

\textsc{Vittorio Del Duca$^a$ and Lorenzo Magnea$^b$}

\vspace{1cm}

$^a$ Institute for Theoretical Physics, ETH Z\"urich, 8093 Z\"urich, 
Switzerland\footnote{On leave from Laboratori Nazionali di Frascati, INFN, Italy.}

\vspace{5mm}

$^b$ Dipartimento di Fisica and Arnold-Regge Center, Universit\`a di Torino, \\
and INFN, Sezione di Torino, Via P. Giuria 1, I-10125 Torino, Italy

\vspace{2cm}

\begin{abstract}
\noindent
The Regge limit of gauge-theory  amplitudes and cross sections is a powerful 
theory tool for the study of fundamental interactions. It is a vast field of research, 
encompassing perturbative and non-perturbative dynamics, and ranging from 
purely theoretical developments to detailed phenomenological applications. 
It traces its origins to the proposal of Tullio Regge, almost sixty years ago, to 
study scattering phenomena in the complex angular momentum plane. 
In this very brief contribution, we look back to the early days of Regge theory, and 
follow a few of the many strands of its development, reaching to present day 
applications to scattering amplitudes in non-abelian gauge theories.
\end{abstract}

\end{center}

\newpage


\section{Introduction}
\label{Intro}

The beginnings of Tullio Regge's career were characterised by deep and early 
forays into new and fast-developing fields of theoretical physics, where his 
extraordinary mastery of mathematical techniques never failed to leave a 
lasting mark. This was the case for his paper with John Wheeler on the stability 
of the Schwarzschild solution in general relativity~\cite{Regge:1957td}, and 
certainly for Regge Calculus~\cite{Regge:1961px}, but it is probably fair to say 
that his most prophetic contribution in those years was his proposal to study 
scattering amplitudes by means of their analiticity properties in the complex 
angular momentum plane~\cite{Regge:1959mz}.

Regge's original work was in the context of non-relativistic potential scattering, 
and it succeeded in uncovering a remarkable connection between the bound states
of a quantum mechanical potential and the behaviour of the corresponding scattering
amplitudes for complex values of the scattering angle. It was however quickly 
realised that these ideas could be generalised to relativistic quantum field 
theory (for a historical account, see the classic textbooks~\cite{Eden:1966dnq,
Collins:1977jy}), where the constraints of Lorentz invariance and crossing 
symmetry would make the formalism even more powerful. In particular, 
singularities in the complex angular momentum plane are tied to the high-energy
behaviour of scattering amplitudes in the crossed channel, which is of
great interest for collider applications.

The generalisation to quantum field theory was spearheaded by Vladimir 
Gribov~\cite{GribovBook}, and the resulting body of knowledge is often
appropriately called {\it Gribov-Regge Theory}: it has turned out to be one 
of the most far-reaching and fruitful set of ideas in high-energy physics in
the past half century. To name just three sets of applications and consequences
of Gribov-Regge theory, we first note that it played a crucial role in the origins
of string theory: indeed, the Veneziano model~\cite{Veneziano:1968yb}
was developed in search of a crossing-symmetric and unitary realization of
Regge's ideas; next, in the context of the theory of strong interactions, at the 
non-perturbative level, linearly rising Regge trajectories provide excellent fits 
to mass spectra (see for example \cite{Ebert:2009ub}), and give simple and 
elegant parametrisations for total hadronic cross sections, across more than 
three orders of magnitude in energy~\cite{Donnachie:1992ny,Donnachie:2013xia};
finally, it has been known since the seminal work of Lev Lipatov~\cite{Lipatov:1976zz}
that Regge behaviour is realised at the perturbative level in the scattering amplitudes
and cross sections of non-abelian gauge theories, in the high-energy limit.
On the one hand, this leads to improved perturbative predictions of great relevance
for collider phenomenology (see for example~\cite{Andersen:2011hs}); on the 
other hand, it provides a laboratory for the most advanced studies of the
properties of scattering amplitudes, in particular in the interesting case of
$N = 4$ Super-Yang-Mills (SYM) theory (see for example~\cite{DelDuca:2009au,
DelDuca:2011ae}).

The above set of examples makes it abundantly clear that any attempt 
to review, even superficially, the developments that flowed from Regge's 
original idea is doomed to fail: literally tens of thousands of papers have 
been written in the various fields of high-energy physics tracing their inspiration 
to Ref.~\cite{Regge:1959mz}. Our goal here will be much more limited: in 
\secn{PotSca} we will briefly illustrate the concept of a `Regge Trajectory',
and we will sketch the implications of Regge's approach for the high-energy 
limit of scattering amplitudes. In \secn{ReggePert}, we will attempt
to provide a bird's eye view of the manifold implications of Regge's ideas
for scattering amplitudes in perturbative non-abelian gauge theories. Finally, 
in \secn{PhenRegge}, we will briefly present three short examples illustrating  
the vast phenomenological relevance of Regge's ideas for the modern 
understanding of strong interactions. Inevitably, our own history and 
viewpoint tilts our account towards the issues that we understand best and 
we have been actively working on: we apologise in advance for the 
omission of many very relevant topics from these notes. We hope that 
nonetheless the remarkable breadth of the impact of Regge Theory on 
contemporary high-energy physics will be apparent.


\section{From potential scattering to field theory}
\label{PotSca}

The basic intuition concerning the connection between poles of the scattering
amplitude in angular momentum and bound states can be gathered by considering
the best-known quantum-mechanical system: the hydrogen atom~\cite{WikiRegge}. 
The textbook solution for energy levels in the Coulomb potential (neglecting spin) is 
\beq
  E \, = \, E_n \, = \, - \frac{\mu e^4}{2 \hbar^2} \, \frac{1}{N^2} \, , \qquad 
  \qquad N \, = \, n + l + 1 \, ,
\label{HydroEn}
\eeq
where $\mu$ is the reduced mass, and the orbital angular momentum quantum 
number $l$ is a non-negative integer. It is clearly possible to invert this expression
and express the angular momentum as a function of the energy $E$. One finds
\beq
  l(E) \, = \, - \, n - 1 + {\rm i} \, \frac{e^2}{\hbar} \, \sqrt{\frac{\mu}{2 E}} \, \equiv  \,
  - \, n + g(E) \, .
\label{HReggeTraj}
\eeq
This is an example of a {\it Regge Trajectory}: as a function of the energy, even
for real $E$, the angular momentum quantum number $l$ sweeps a curve in 
the complex plane, crossing the real axis at non-negative integer points
when the energy equals a bound state energy $E_n$. For the Coulomb 
problem, one can also compute the $S$-matrix as a function of energy and 
angular momentum. The answer can be written as~\cite{Weinberg}
\beq
  S (E, l) \, = \, \frac{\Gamma \left[ l - g(E) \right]}{\Gamma \left[ l + g(E) \right]} \, \, 
  {\rm e}^{{\rm i} \pi l} ~.
\label{CoulS}
\eeq
As expected, the $S$-matrix has simple poles, due to the numerator $\Gamma$ 
function, where $l$ and $E$ satisfy the bound state quantisation condition.

In order to generalise this observation to other potentials, and ultimately to relativistic
field theory, a good starting point is the well-known partial-wave expansion of the
scattering amplitude in a series of Legendre polynomials. For a 2-particle scattering
process we write
\beq
  A(s, \theta) \, = \, 16 \pi  \sum_{l = 0}^\infty \left(2 l + 1 \right) a_l (s) P_l 
  \left( \cos \, \theta \right) \, ,
\label{partwa}
\eeq
where $s$ is squared center-of-mass energy and $\theta$ the scattering angle in
the center-of-mass frame in the $s$-channel. For equal masses, the scattering angle 
$\theta$ is related to the squared momentum transfer $t$ by
\beq
  t \, = \, - \, \frac{1}{2} \left(s - 4 m^2 \right) \left( 1 - \cos \theta \right) \, ,
\label{tang}
\eeq
so that $t \leq 0$ above threshold. \eq{partwa} can be inverted to yield the partial-wave 
amplitude $a_l(s)$ according to
\beq
  a_l (s) \, = \, \frac{1}{2} \int_{-1}^1 d z \, A (s, z) \, P_l (z) \, ,
\label{partwa2}
\eeq
The crucial step taken by Regge~\cite{Regge:1959mz,Bottino:1962aqa} was to 
propose that the partial-wave amplitude in \eq{partwa2}, defined for integer $l$, 
could be usefully generalised to an analytic function $a(l, s)$ depending on the 
complex variables $l$ and $s$. Under suitable conditions on the asymptotic 
behaviour for large $l$, which can be shown to be met for potential scattering, 
the definition of the `interpolating function' $a(l,s)$ is unique. One can then 
derive an expression for the scattering amplitude in terms of a contour integral 
in the $l$ plane, performing a Sommerfeld-Watson transform\footnote{In this 
brief account of the consequences of treating $l$ as a complex variable we 
sketchily follow Ref.~\cite{Eden:1966dnq}. For a recent historical perspective, 
see also~\cite{Bottino:2018zrc}.}. For example, assume $a(l,s)$ to be analytic 
in $l$ for ${\bf Re} \, l > l_0$ and not growing too fast for $l \to \infty$, and let 
$n$ be the smallest integer $n > l_0$: then one can write
\beq
  A(s, \theta) \, = \, 16 \pi  \sum_{l = 0}^{n - 1} \left(2 l + 1 \right) a_l (s) P_l 
  \left( \cos \, \theta \right) \, + \, 8 \pi {\rm i} \int_{\cal C} d l \, \frac{(2 l + 1) \, a(l,s) 
  \, P_l ( - \cos \, \theta )}{\sin(\pi l)} \, ,
\label{SomWat}
\eeq
with the integration contour ${\cal C}$ closely encircling the real axis for $l > l_0$, 
enclosing the poles induced by the denominator for ${\bf Re}\, l \geq n$. One may 
now attempt to open up the integration contour, making it parallel to the imaginary 
axis, and then move it to smaller values of ${\bf Re} \, l$. In the absence of complex
singularities in $a(l,s)$, this just reconstructs the original partial-wave expansion; 
if however $a(l,s)$ has, for example, isolated poles in the $l$ plane away from 
the real axis, in the process of moving the integration contour one picks up the 
corresponding residues. The location of these poles will in general depend on $s$, 
so that, varying $s$, the poles will follow trajectories such as \eq{HReggeTraj}. 
Denoting the trajectory of the $i$-th pole by $l_i = \alpha_i (s)$, and the residue 
of that pole by $r_i(s)$, the contribution of the poles to the scattering amplitude 
will be proportional to
\beq
  \sum_i \frac{ \left( 2 \alpha_i (s) + 1 \right) \, r_i (s) \, P_{\alpha_i (s)} 
  ( - \cos \theta )}{\sin \left( \pi \alpha_i (s) \right)} \, .
\label{polecontr}
\eeq
One can now use the knowledge of the asymptotic behaviour of Legendre 
polynomials
\beq
  P_l (- z) \, \sim  \, \left( - z \right)^l \, , \qquad  z \to \infty \,  ,
\label{aspl}
\eeq
to prove that in the (unphysical) limit $\cos \theta \to \infty$ (corresponding 
to $t \to \infty$ by \eq{tang}) the scattering amplitude is dominated by the 
contribution of the rightmost pole in the angular momentum plane, $l_{\rm max}
= \alpha_{\rm max} (s)$, and behaves like an $s$-dependent power of $t$ for
large $t$.

Even though the large $t$ region is not physical, the result just derived is
of great interest: since Regge trajectories are related to bound states for the 
potential being considered, we now have a relation between the location of
bound states and the asymptotic behaviour of the scattering amplitude; this
allows, for example, to broadly extend the domain of applicability of dispersion 
relations. The extension of this result to relativistic field theory is however
much more powerful: by Lorentz invariance and crossing symmetry, one 
can in fact transfer the $s$-channel argument to the $t$ channel, and derive
results for the asymptotic behaviour of the scattering amplitude for large 
$s$, which is physical, and of great phenomenological interest.

The main tool for achieving the relativistic generalisation is the use of dispersion
relations to reconstruct the scattering amplitude in terms of its discontinuities
along the cuts originating from normal thresholds. For example, for equal mass
particles and up to rational contributions, one may write
\beq
  A(s,t) \, = \, \frac{1}{\pi} \left[ \int_{4 m^2}^\infty d t' \, \frac{ {\rm Disc}_t \!
  \left( A(s,t) \right)}{t' - t} + \int_{4 m^2}^\infty d u' \, \frac{ {\rm Disc}_u \! 
  \left( A(s,t) \right)}{u' - u} \right] \, .
\label{Disprel}
\eeq
Using \eq{Disprel} in \eq{partwa2}, one may perform the $z$ integration by means
of Neumann's formula
\beq
  \frac{1}{2} \int_{-1}^1 d w \, \frac{P_l (w)}{z - w} \, = \, Q_l (z) \, ,
\label{leg2}
\eeq
where $Q_l (z)$ are the Legendre functions of the second kind. This leads to the
Froissart-Gribov representation of the partial wave amplitude
\beq
  a_l(s) \, = \, \frac{1}{\pi} \left[ \int_{z_s}^\infty d z \, {\rm Disc}_t \!
  \left[ A(s, t(z)) \right] \, Q_l (z) - \int_{z_s}^\infty d z \, {\rm Disc}_u \! 
  \left[ A( s, t(z) ) \right] \,  Q_l(-z) \right] \, ,
\label{FroGri}
\eeq
where $z_s = 1 + 2 m^2/s$ in the equal-mass case. The Sommerfeld-Watson 
transform is not directly applicable to \eq{FroGri} because of the bad convergence 
properties of the second term for large $l$. This leads to the introduction of the 
concept of {\it signature}: one needs to consider the (anti)symmetric combinations
of $t$- and $u$-channel contributions defined by
\beq
  a_l^\pm (s) \, = \, \frac{1}{\pi} \int_{z_s}^\infty d z \Big[ {\rm Disc}_t \!
  \left[ A(s, t(z)) \right] \, \pm \, {\rm Disc}_u \! \left[ A( s, t(z) ) \right] \Big] \, Q_l (z) \, .
\label{Signa}
\eeq
Applying Regge's reasoning to partial waves of definite signature constrains
the asymptotic behaviour of relativistic scattering amplitudes. In particular, and 
most interestingly, one can use crossing symmetry to study partial waves in the
$t$-channel physical region, and for complex $l$: the presence of a `Regge' 
pole in these partial wave amplitudes leads to a $t$-dependent power behaviour
of the scattering amplitude at high energy, according to
\beq
  a_l^\pm (t) \, \simeq \, \frac{1}{l - \alpha (t)}  \qquad \longrightarrow \qquad
  A(s,t) \, \stackrel{s \to \infty}{\longrightarrow} \, f(t) \, s^{\alpha(t)} \, .
\label{Repol}
\eeq
Clearly, \eq{Repol} is only the simplest example of connection between singularities
for complex $l$ and asymptotic behaviours. Indeed, it was clear from the early days 
that one should in general expect contributions from `Regge cuts', which are required
at the non-perturbative level by unitarity arguments, and can be seen to arise 
perturbatively from the analysis of high-order Feynman diagrams. In the decades
following Regge's original contribution, the manifold consequences of these subtle
analytic properties for scattering amplitudes have been steadily explored, with
important consequences both for non-perturbative studies of strong interactions,
and for the exploration of perturbative gauge-theory scattering amplitudes.


\section{The Regge limit for perturbative  gauge theories}
\label{ReggePert}

The discussion in Section~\ref{PotSca} is general, and, for quantum field theories,
it is based only on Lorentz invariance and analiticity: as a consequence, it is expected
to apply in both the perturbative and non-perturbative regimes. When it became
clear that strong interactions are described by QCD, and that the theory admits a 
rich and very relevant perturbative sector, a natural question was how Regge 
theory results would emerge from the perturbative expansion of a non-abelian
gauge theory.

\begin{figure}
  \centerline{\includegraphics[width=0.3\columnwidth]{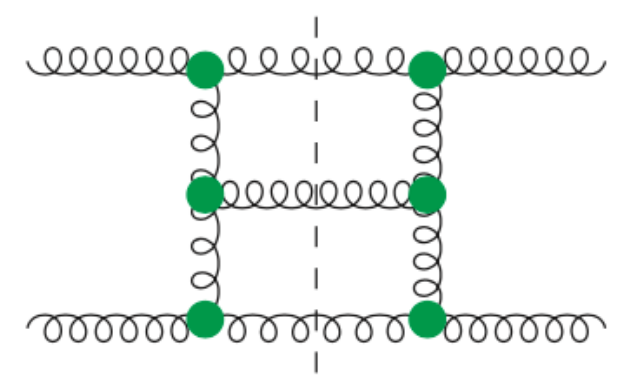} ~~~~~~
  \includegraphics[width=0.29\columnwidth]{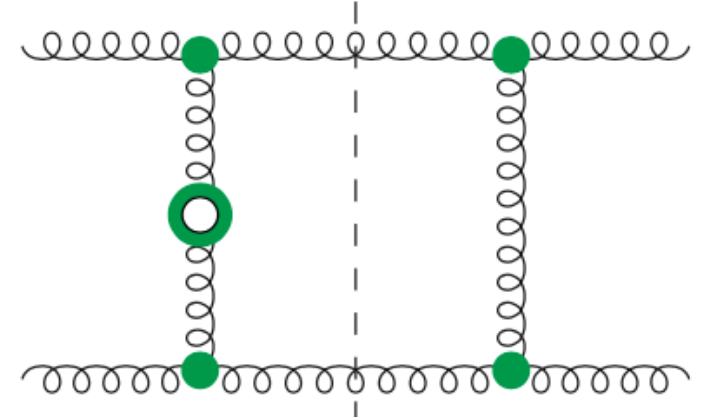} }
  \caption{$(a)$ Emission of a gluon along the ladder; $(b)$ One-loop 
  Regge trajectory. } 
\label{fig:llbfkl}
\end{figure}

In the Regge limit, in which the squared centre-of-mass energy $s$ is much 
larger than the momentum transfer $|t|$, $s\gg |t|$, any scattering process 
is dominated by the exchange in the $t$ channel of the highest-spin particle. 
In the case of QCD, or ${\cal N}=4$ Super Yang-Mills (SYM), that entails the 
exchange of a gluon in the $t$ channel. The study of this limit for QCD was 
pioneered by Lev Lipatov~\cite{Lipatov:1976zz}, and the results are described 
by the Balitsky-Fadin-Kuraev-Lipatov (BFKL) theory, which models strong-interaction 
processes with two large and disparate scales ($s$ and $t$) by resumming 
radiative corrections to parton-parton scattering. This is achieved, at leading 
logarithmic (LL) accuracy in $\ln(s/|t|)$, through the BFKL equation~\cite{Fadin:1975cb,
Kuraev:1976ge,Kuraev:1977fs,Balitsky:1978ic}, an integral equation for the 
evolution of $t$-channel gluon exchange in transverse-momentum space 
and Mellin moment space, which can be written as 
\beq
  \omega\, f_\omega(q_1, q_2) \, = \, {1\over 2} \, \delta^{(2)} (q_1 - q_2) + 
  \left( \cK \star f_\omega \right) (q_1,q_2) \, .
\label{bfklop}
\eeq
In \eq{bfklop}, $f_\omega$ is related by Laplace-Mellin transform to the discontinuity
of the four-point amplitude in the Regge limit, and depends on the momenta flowing
in the $t$-channel on either side of the cut; the convolution is defined by
\beq
  \left( \cK \star f_\omega \right) (q_1,q_2) \, \equiv \, \int d^2 k \, K (q_1, k) \,
  f_{\omega} (k, q_2) \, ,
\label{convo}
\eeq
and $K(q_1,q_2)$ the BFKL kernel. The kernel is real and symmetric, 
$K(q_1,q_2) = K(q_2,q_1)$, so that the integral operator $\cK$ is hermitian 
and its eigenvalues are real. At LL accuracy in $\ln(s/|t|)$, the building blocks 
of the BFKL kernel are a real correction (the emission of a gluon along the 
ladder, {\it a.k.a.} the central emission vertex, displayed in Fig.~\ref{fig:llbfkl}$(a)$),
and a virtual correction (the one-loop Regge trajectory~\cite{Lipatov:1976zz}, 
displayed in Fig.~\ref{fig:llbfkl}$(b)$). The BFKL equation is then obtained by 
iterating those one-loop corrections to all orders in the strong coupling $\alpha_s$,
resumming the terms of ${\cal O}(\alpha_s^n \ln^n(s/|t|))$. 

Underpinning BFKL theory is the diagrammatic ladder structure of scattering 
amplitudes in the Regge limit. For instance, in the case of gluon-gluon scattering 
in QCD, the gluon ladder is described by the $8 \otimes 8$ colour representation, 
which is decomposed as $8 \otimes 8 = \{1 \oplus 8_s \oplus 27\} \oplus [8_a 
\oplus 10 \oplus \bar{10}]$, where the term in curly brackets in the direct sum 
is even under $s \leftrightarrow u$ exchange, while the term in square brackets 
is odd. To LL accuracy in $\ln(s/|t|)$, however, parton-parton scattering amplitudes
are real and only the antisymmetric octet contributes: this implies gluon 
Reggeisation at LL accuracy~\cite{Lipatov:1976zz}. For phenomenological 
purposes, the solution of the BFKL equation must then to be endowed with 
process-dependent impact factors describing the evolution of the emitting partons: 
the simplest cases are the quark and gluon impact factors~\cite{Kuraev:1976ge,
DelDuca:1995zy}, by means of which one can obtain, for example, the di-jet 
cross section at large rapidities~\cite{Mueller:1986ey}.

The kernel of the BFKL equation (\ref{bfklop}) admits the perturbative 
expansion,
\beq
  K(q_1,q_2) \, = \, \overline{\alpha}_\mu \sum_{l = 0}^\infty
  \overline{\alpha}_\mu^l \, K^{(l)} (q_1,q_2) \, .
\label{pertkern}
\eeq
where $\overline{\alpha}_\mu = N_C \, \alpha_s(\mu^2)/\pi$ is the renormalised 
strong coupling constant evaluated at an arbitrary scale $\mu^2$. $K^{(0)}$ is 
the leading-order BFKL kernel, which leads to the resummation of terms of 
${\cal O}(\alpha_s^n \ln^n(s/|t|))$, corresponding to the LL accuracy discussed 
above; similarly, the NLO kernel $K^{(1)}$ resums terms at next-to-leading-logarithmic
(NLL) accuracy, {\it i.e.} of ${\cal O}(\alpha_s^n \ln^{n - 1}(s/|t|))$, and so forth. 

\begin{figure}
  \centerline{\includegraphics[width=0.25\columnwidth]{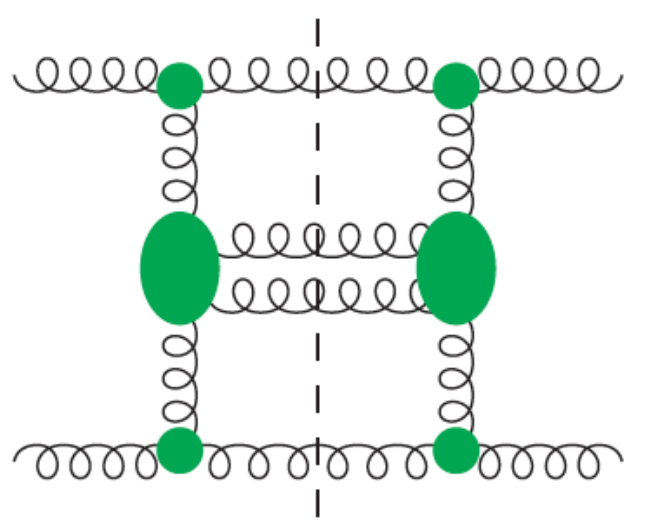} ~~
  \includegraphics[width=0.25\columnwidth]{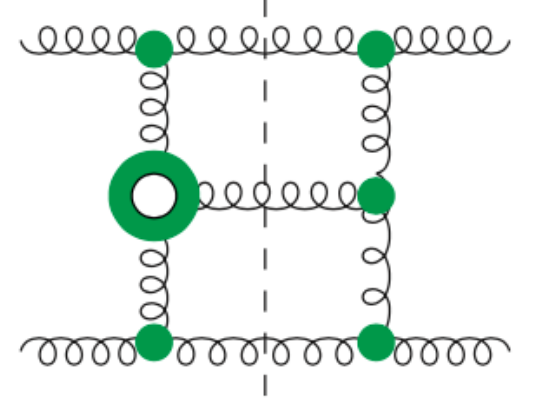} ~~
  \includegraphics[width=0.31\columnwidth]{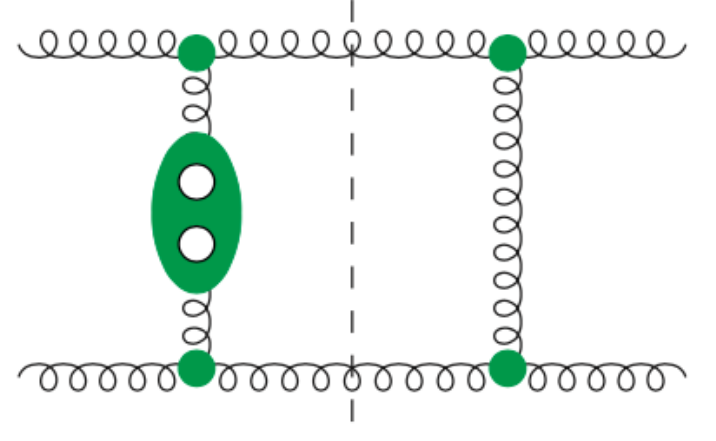} }
  \caption{$(a)$ Emission of two gluons (or of a quark-antiquark pair) along 
  the ladder; $(b)$ one-loop corrections to the central emission vertex;
  $(c)$ two-loop Regge trajectory. } \label{fig:nllbfkl}
\end{figure}

The computation of the NLO BFKL kernel~\cite{Fadin:1998py,Ciafaloni:1998gs,
Kotikov:2000pm,Kotikov:2002ab} requires the real corrections to the central emission 
vertex, corresponding to the emission of two gluons, or of a quark-antiquark pair, 
along the ladder, as depicted in Fig.~\ref{fig:nllbfkl}$(a)$~\cite{Fadin:1989kf,
DelDuca:1995ki,Fadin:1996nw,DelDuca:1996nom}; furthermore, one needs 
the one-loop correction to the central emission vertex (Fig.~\ref{fig:nllbfkl}$(b)$)
\cite{Fadin:1993wh,Fadin:1994fj,Fadin:1996yv,DelDuca:1998cx,Bern:1998sc},
as well as the two-loop Regge trajectory (Fig.~\ref{fig:nllbfkl}$(c)$)~\cite{Fadin:1995xg,
Fadin:1996tb,Fadin:1995km,Blumlein:1998ib,DelDuca:2001gu}. Underpinning 
BFKL theory at NLL accuracy is the fact that, while parton-parton scattering 
amplitudes develop an imaginary part at one loop, the BFKL equation is computed 
through the real part of the amplitudes, which receives contributions only from 
the antisymmetric octet representation; this, again, implies gluon Reggeisation 
at NLL accuracy~\cite{Fadin:2006bj,Fadin:2015zea}.

At LL accuracy, the kernel of the BFKL equation is conformally invariant, and thus 
the leading-order eigenfunctions of the BFKL equation are fixed~\cite{Lipatov:1985uk}  
by conformal symmetry. NLL corrections~\cite{Fadin:1998py,Ciafaloni:1998gs,
Kotikov:2000pm,Kotikov:2002ab} to the BFKL equation were computed by 
acting with the next-to-leading-order (NLO) kernel of the equation on the 
leading-order eigenfunctions. This procedure is not fully consistent, and it 
was already clear to Fadin and Lipatov~\cite{Fadin:1998py} that the terms 
which make the procedure inconsistent are related to the running of the 
coupling. Consistent NLO eigenfunctions were later constructed by Chirilli and 
Kovchegov~\cite{Chirilli:2013kca,Chirilli:2014dcb}, who found that indeed the 
additional pieces which occur at NLO are proportional to the QCD beta function.


\subsection{The Regge limit and infrared factorisation}
\label{IRRegge}

The backbone of the BFKL equation are the quark and gluon scattering 
amplitudes, which at high energy are dominated by the exchange of a gluon 
in the $t$ channel. Because loop-level scattering amplitudes of massless 
partons are infrared divergent, and so are the impact factors, the Regge 
trajectory, and the central emission vertex, which build up the amplitudes 
in the Regge limit, the study of scattering amplitudes at high energy has 
benefited from a cross breeding with infrared factorisation~\cite{Sotiropoulos:1993rd,
Korchemsky:1993hr,Korchemskaya:1994qp,Korchemskaya:1996je,Bret:2011xm,
DelDuca:2011ae,Caron-Huot:2013fea,DelDuca:2013ara,DelDuca:2014cya,
Caron-Huot:2017fxr,Caron-Huot:2017zfo}. In particular infrared factorisation
leads to the exponentiation of infrared singularites: in the high-energy limit,
this leads to all-order expressions for singular contributions to Regge-theory
quantities such as the Regge trajectory and the impact factors.

Infrared (soft and collinear) divergences in a multi-parton amplitude ${\cal A}$ 
are known to factorise according to
\beq
  {\cal A} \left( \frac{p_i}{\mu}, \alpha_s (\mu), \epsilon \right) \, = \, 
  {\bf Z} \left( \frac{p_i}{\mu}, \alpha_s(\mu), \epsilon \right) \, {\cal H}
  \left( \frac{p_i}{\mu}, \alpha_s(\mu), \epsilon \right) \, ,
\label{IRfac}
\eeq
where the scattering amplitude ${\cal A}$ and the finite hard factor ${\cal H}$ 
should be seen as vectors in the space of available color configurations, 
while IR divergences are generated by the color operator ${\bf Z}$. Using 
renormalisation-group arguments, ${\bf Z}$ can be expressed in terms of
an anomalous dimension matrix, known up to three loops in the massless 
case~\cite{Aybat:2006wq,Aybat:2006mz,Dixon:2008gr,Becher:2009cu,
Gardi:2009qi,Gardi:2009zv,Becher:2009qa,Dixon:2009ur,Almelid:2015jia,
Almelid:2017qju}. In the Regge limit, and at least to three-loop accuracy, 
the infrared-divergent operator ${\bf Z}$ simplifies considerably, and takes 
the form~\cite{DelDuca:2011ae,DelDuca:2014cya,Caron-Huot:2017fxr}
\beq
  {\bf Z} \left( \frac{p_i}{\mu}, \alpha_s(\mu), \epsilon \right) \, = \, 
  \widetilde{{\bf Z}} \left( \frac{s}{t}, \alpha_s(\mu), \epsilon \right)
  Z_i \left( \frac{t}{\mu^2}, \alpha_s(\mu), \epsilon \right) \, 
  Z_j \left( \frac{t}{\mu^2}, \alpha_s(\mu), \epsilon \right) \,,
\label{factZ}
\eeq
where the indices $i$ and $j$ identify the high-energy partons in the selected 
channel, and the factors $Z_i$ and $Z_j$ are color-singlet quantities given by
\beq
  Z_i \left( \frac{t}{\mu^2}, \alpha_s(\mu), \epsilon \right) = \, 
  {\rm P} \exp \left\{ - \int_0^{\mu^2} \frac{d \lambda^2}{\lambda^2}
  \left[ \frac{\gamma_K \left( \alpha_s(\lambda) \right)}{4} \,
  \ln \frac{- t}{\lambda^2} \right] + \gamma_i \left( \alpha_s(\lambda) \right)
  \right\} \, .
\label{Zscal}
\eeq
\eq{Zscal} naturally defines infrared divergent contributions to the impact 
factors in terms of the cusp $\gamma_K$ and collinear $\gamma_i$ 
anomalous dimensions. Colour and energy dependences, on the other 
hand, are confined to the operator $\widetilde{{\bf Z}}$, defined by
\beq
  \widetilde{{\bf Z}} \left( \frac{s}{t}, \alpha_s(\mu), \epsilon \right)
  = \exp \left\{ \! K \big( \alpha_s(\mu), \epsilon \big) \! \left[
  \left( \ln \left( \frac{s}{- t} \right) - {\rm i} \frac{\pi}{2} \right)
  {\bf T}_t^2 + {\rm i} \pi {\bf T}_{s - u}^2 \right] + Q_{\bf \Delta}^{(3)}
  \big( \alpha_s(\mu), \epsilon \big) \right\} .
\label{Ztil}
\eeq
In \eq{Ztil} we have introduced the colour operators
\beq
  {\bf T}_s \, = \, {\bf T}_1 + {\bf T}_2 \, , \quad 
  {\bf T}_u \, = \, {\bf T}_1 + {\bf T}_3 \, , \quad 
  {\bf T}_t \, = \, {\bf T}_1 + {\bf T}_4 \, , \quad 
  {\bf T}_{s - u}^2 \, = \, \frac{1}{2} \big( {\bf T}_s^2 - {\bf T}_u^2 \big) \, ,
\eeq
where ${\bf T}_i$ is a gluon insertion operator acting on parton $i$. Up to two
loops, the exponent in \eq{Ztil} is completely determined by dipole colour 
correlations, and all singularities are generated by the integral
\beq
  K \big( \as (\mu), \epsilon \big) \, = \, - \frac{1}{4} \int_0^{\mu^2} 
  \frac{d \lambda^2}{\lambda^2} \,
  \widehat{\gamma}_K \left( \alpha_s (\lambda) \right) \, ,
\label{cusp}
\eeq
which relates the infrared-divergent part of the Regge trajectory to the cusp 
anomalous dimension\footnote{In \eq{cusp}, $\widehat{\gamma}_K(\alpha_s)$ 
is the cusp anomalous dimension, rescaled by the quadratic Casimir eigenvalue 
of the relevant parton representation.}, to all orders in perturbation theory. 
At three loops, quadrupole contributions arise for the first time, and, in the 
high-energy, limit, they contribute through the integral
\beq
  Q_{\bf \Delta}^{(3)} \big( \as (\mu), \epsilon \big) \, = \, - \frac{{\bf \Delta}^{(3)}}{2} 
  \int_0^{\mu^2} \frac{d\lambda^2}{\lambda^2}
  \left( \frac{\alpha_s(\lambda)}{\pi} \right)^3 \, ,
\label{threelo}
\eea
where ${\bf \Delta}^{(3)}$ is a single-logarithmic contribution with an intricate 
colour structure~\cite{Caron-Huot:2017fxr}.

Using Eqs.~(\ref{factZ}-\ref{threelo}) one can, for example, determine the 
singular contributions to one-loop impact factors in terms of the one-loop cusp 
anomalous dimension and the one-loop quark and gluon collinear anomalous 
dimensions~\cite{DelDuca:2013ara,DelDuca:2014cya}; similarly, singular 
contributions to the two-loop trajectory emerge from the two-loop cusp 
anomalous dimension~\cite{Korchemskaya:1996je}. Furthermore, we 
know that the picture of Regge-pole factorisation, based on gluon 
Reggeisation, breaks down at NNLO accuracy~\cite{DelDuca:2001gu}.
The origin of the violation can be explained by means of infrared factorisation, 
by showing that the real part of the amplitudes becomes non-diagonal in the 
$t$-channel-exchange basis~\cite{Bret:2011xm,DelDuca:2011ae}. Accordingly, 
it is possible to predict how the violation propagates to higher loops, and the 
three-loop prediction~\cite{DelDuca:2013ara,DelDuca:2014cya}, which has 
NNLL accuracy, has been confirmed by the explicit computation of the 
three-loop four-point function of full ${\cal N}=4$ SYM in Ref.~\cite{Henn:2016jdu}. 
In the Regge-pole factorisation picture, the violation is due to the exchange 
of three Reggeised gluons~\cite{Fadin:2016wso,Caron-Huot:2017fxr}. 
Thus, although a full study of the BFKL ladder at NNLL accuracy has 
yet to be undertaken (the three-loop Regge trajectory in a specific 
scheme~\cite{Caron-Huot:2017fxr}, and the emission of three partons 
along the gluon ladder being the only building blocks computed so 
far~\cite{DelDuca:1999iql,Antonov:2004hh,Duhr:2009uxa}), we already 
have a clear picture of how the factorisation violations occur at NNLL 
accuracy.

While we have a precise knowledge of how infrared poles occur in loop 
corrections to impact factors and to the Regge trajectory, the finite parts of 
those corrections are not constrained by either Regge or infrared factorisation, 
and must be directly computed. Recently, however, a relation has been 
exposed~\cite{DelDuca:2017pmn} between the ${\cal O}(\epsilon)$ terms of 
the one-loop gluon impact factor~\cite{Bern:1998sc} and the ${\cal O}(\epsilon^0)$ 
terms of the two-loop Regge trajectory~\cite{Fadin:1995xg,Fadin:1996tb,
Fadin:1995km,Blumlein:1998ib,DelDuca:2001gu}: this hints at a connection 
between the Regge limit and infrared factorisation extending beyond the infrared 
poles.


\subsection{The Regge limit in ${\cal N}=4$ SYM and the analytic structure 
of scattering amplitudes}
\label{SYMRegge}

In the last few years, it has been realised that the Regge limit of QCD and of 
${\cal N}=4$ SYM, and thus the BFKL equation, are endowed with a rich 
mathematical structure: the Regge limit has thus become a laboratory for
advanced studies of gauge-theory scattering amplitudes at high orders.

It is well-known that ${\cal N} = 4$ SYM is a super-conformal theory. In the 
planar limit, the scattering amplitudes of the theory also exhibit a dual 
conformal symmetry~\cite{Drummond:2006rz,Alday:2007hr,Drummond:2007aua,
Brandhuber:2007yx}, where the conformal group acts on dual variables $x_i$
defined in terms of (cyclically ordered) particle momenta by
\beq
  p_i \, = \, x_i - x_{i - 1} \, .
\label{dualvar}
\eeq
Dual conformal invariance is broken by infrared divergences~\cite{Drummond:2007cf,
Drummond:2007au}, but it is possible to construct finite functions of dual-conformal 
invariant ratios of $x_i$'s, in which all infrared divergences cancel. As a 
consequence, the analytic structure of scattering amplitudes in planar ${\cal N} = 4$
SYM is highly constrained. In particular, the four and five-point amplitudes are 
fixed to all loop orders by their symmetries, in terms of the one-loop amplitudes 
and of the cusp anomalous dimension~\cite{Drummond:2007au,Bern:2005iz}. 
The first occurrence of non-trivial corrections is then for amplitudes with at 
least six legs~\cite{Drummond:2007au,Bern:2008ap,Drummond:2008aq,
DelDuca:2009au,DelDuca:2010zg}. The ordinary and dual conformal symmetries 
are also at the heart of a surprising identity between scattering amplitudes and 
Wilson loops computed along light-like polygonal contours~\cite{Alday:2007hr,
Drummond:2007aua,Brandhuber:2007yx,Drummond:2007au,Drummond:2007cf,
Drummond:2008aq} in the dual space parametrised by the $x_i$ coordinates. 
Thus, in what follows, any statement about amplitudes applies also to the 
corresponding polygonal Wilson loops.

The relevance of the Regge limit for studies of these amplitudes became 
apparent when it was observed that the four-point amplitude is Regge 
exact~\cite{Drummond:2007aua,Naculich:2007ub}, {\it i.e.} it is not 
modified by the Regge limit. Regge exactness extends to the five-point 
amplitude~\cite{Brower:2008nm,Bartels:2008ce}, if computed in multi-Regge 
kinematics (MRK)~\cite{Lipatov:1976zz}, characterised by the central-emission 
vertex; similarly, it extends to the six-point amplitude~\cite{DelDuca:2009au}, 
if computed in quasi-multi-Regge kinematics~\cite{Fadin:1989kf},
characterised by the emission of two gluons along the ladder. In fact, 
Regge exactness extends in general to $n$-point MHV amplitudes,
if computed in quasi-multi-Regge kinematics for the emission of $(n - 4)$ 
gluons along the ladder~\cite{DelDuca:2010zg}.

\begin{figure}
  \centerline{\includegraphics[width=0.25\columnwidth]{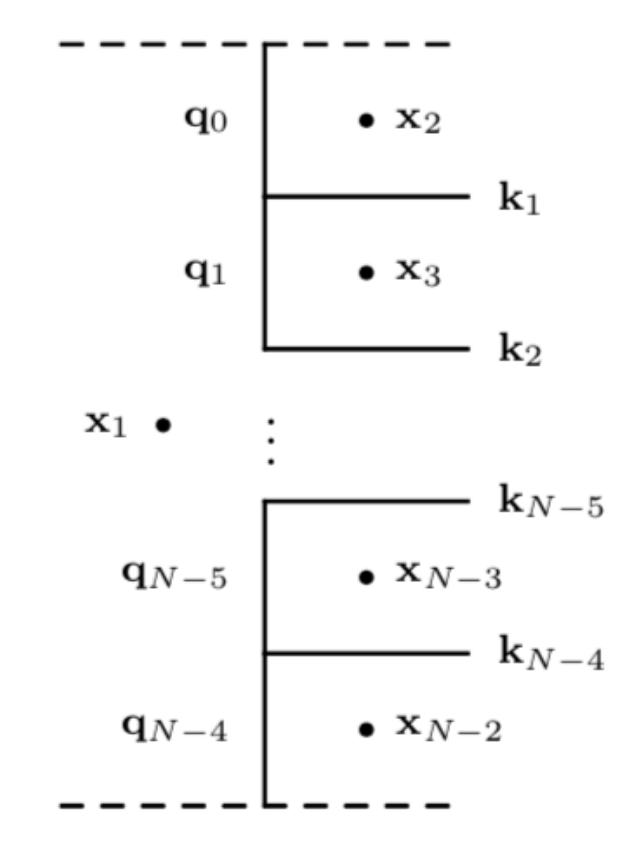}  }
  \caption{Dual coordinates in transverse space for the ladder of the
  $n$-point amplitude. } 
\label{fig:n-5}
\end{figure}

The Regge limit provides further insights in the mathematical structure of 
amplitudes in planar ${\cal N}=4$ SYM, specifically on the functional spaces 
where the amplitudes are defined. In general, it is expected that $n$-point 
amplitudes in this theory should be expressed as iterated integrals of 
differential one-forms~\cite{Chen:1977oja} defined on the space of 
configurations of points in three-dimensional projective space, $\textrm{Conf}_N
(\mathbb{CP}^3)$~\cite{Golden:2013xva,Golden:2014xqa}. The simplest 
instance of iterated integrals that one encounters in these amplitudes are 
multiple polylogarithms (MPL)~\cite{Goncharov:2001iea,Brown:2009qja}, 
which correspond to iterated integrals over rational functions. It is believed 
that all maximally helicity violating (MHV) and next-to-MHV amplitudes in 
planar ${\cal N}=4$ SYM should be expressed in terms of MPLs having uniform 
transcendental weight~\cite{ArkaniHamed:2012nw}. The Regge limit 
provides evidence and abundant data to test these ideas. In fact, in the 
Euclidean region, where all Mandelstam invariants are negative, scattering 
amplitudes in planar ${\cal N}=4$ SYM in multi-Regge kinematics factorise to 
all orders in perturbation theory into building blocks: the impact factor, the 
central-emission vertex, and the Regge trajectory, describing the Reggeised 
gluons exchanged in the $t$-channel, and the emission of gluons along the 
$t$-channel ladder. The building blocks are determined to all orders by the 
four and five-point amplitudes, hence scattering amplitudes in multi-Regge 
kinematics are trivial in the Euclidean region~\cite{Brower:2008nm,
Bartels:2008ce,DelDuca:2008pj,Bartels:2008sc,Brower:2008ia,DelDuca:2008jg}. 
In particular, the Regge trajectory is fixed, to all orders in perturbation theory, 
by the cusp anomalous dimension and the gluon collinear anomalous 
dimension~\cite{DelDuca:2008pj}. Starting from six legs, however, scattering 
amplitudes exhibit Regge cuts, if the multi-Regge limit is taken after analytic 
continuation to a specific Minkowski region~\cite{Bartels:2008ce,Bartels:2008sc}. 
The discontinuity across these cuts is described, to all orders, by a dispersion 
relation closely related to the BFKL equation, which can be expressed in 
terms of single-valued functions~\cite{Lipatov:2010ad,Fadin:2011we}. In 
particular, the discontinuity of the six-point amplitude~\cite{Dixon:2012yy} 
in multi-Regge kinematics can be expressed in terms of single-valued 
harmonic polylogarithms (SVHPL)~\cite{BrownSVHPLs}. Further, it was later
realised that $n$-point amplitudes in multi-Regge kinematics are described 
by the geometry of the moduli space $\mathfrak{M}_{0,n-2}$ of Riemann 
spheres with $(n-2)$ marked points~\cite{DelDuca:2016lad}. It can be shown 
that the algebra of iterated integrals on $\mathfrak{M}_{0,n}$ factors through 
certain hyperlogarithm algebras~\cite{Brown:2009qja}, so that the iterated 
integrals on $\mathfrak{M}_{0,n}$ can always be expressed in terms of 
MPLs and rational functions, with singularities when two marked points 
coincide. To be more precise, one can introduce dual coordinates in 
transverse space, according to
\beq
  {\bf q}_i \, = \, {\bf x}_{i + 2} - {\bf x}_1 \, , \qquad \quad
  {\bf k}_i \, = \, {\bf x}_{i + 2} - {\bf x}_{i + 1} \, ,
\label{trasnco}
\eeq
as illustrated in Fig~\ref{fig:n-5}. The discontinuity of the $n$-point amplitude 
in multi-Regge kinematics is then parametrised by $(n - 5)$ conformally invariant 
cross-ratios, given by
\beq
  z_i \, = \, \frac{ ( {\bf x}_1 - {\bf x}_{i + 3} ) ( {\bf x}_{i + 2} - {\bf x}_{i + 1} ) }{ 
  ( {\bf x}_1 - {\bf x}_{i + 1} ) ( {\bf x}_{i + 2} - {\bf x}_{i + 3} ) } \, = \,  
  - \frac{{\bf q}_{i + 1} {\bf k}_i}{{\bf q}_{i - 1} {\bf k_{i + 1}} }
  \qquad i = 1, \ldots , N - 5 \, .
\label{cicrs}
\eeq
It can be shown~\cite{DelDuca:2016lad} that, in the soft limit for any of the
$(n-5)$ transverse momenta of the gluons emitted along the ladder, one 
of the $(n - 5)$ ratios $z_i$ vanishes, and thus two marked points on the 
Riemann sphere coincide. The transverse momenta of the ladder gluons,
however, never vanish. This requirement constrains the iterated integrals 
that can appear in multi-Regge kinematics to be single-valued functions:
more precisely, linear combinations of products of iterated integrals 
on $\mathfrak{M}_{0,n}$ (and their complex conjugates) such that all 
branch cuts cancel. These are the single-valued multiple polylogarithms 
(SVMPL)~\cite{BrownSVHPLs,BrownSVMPLs,Brown:2013gia}. The 
statements of Ref.~\cite{DelDuca:2016lad}, which were made at LL 
accuracy, can be generalised at NLL accuracy and beyond~\cite{DelDuca:2018hrv,
Marzucca:2018ydt}.

The foundation of the analysis of $n$-point amplitudes in planar 
${\cal N}=4$ SYM in multi-Regge kinematics is provided by Lipatov's 
picture~\cite{Lipatov:2009nt} of the BFKL-like dispersion relation, which 
describes the discontinuity of the amplitude in terms of the Hamiltonian of 
an integrable open Heisenberg spin chain~\cite{Lipatov:1993yb,Faddeev:1994zg}. 
According to this picture, the discontinuity of the amplitude in multi-Regge 
kinematics and in specific Minkowski regions is described by a two-Reggeon 
cut; similarly, the double discontinuity is described by a three-Reggeon cut, 
and in general the $m$-fold discontinuity is described by an $(m+1)$-Reggeon 
cut. As outlined above, this picture has been fully explored at the level of the 
single discontinuity, while the analysis of the double discontinuity has just 
begun~\cite{DelDuca:2018raq}.

Previously, Lipatov had characterised the BFKL equation, and, more in general,
the exchange of a ladder of $n$ Reggeised gluons in a singlet configuration 
in QCD at large $N_c$, in terms of the Hamiltonian~\cite{Lipatov:1993yb,
Lipatov:1993qn,Lipatov:1998as} of a closed spin chain. It should then come 
as no surprise that, just like the case of the six-point amplitude of ${\cal N}=4$ 
SYM in multi-Regge kinematics~\cite{Dixon:2012yy}, also the analytic 
structure of the BFKL ladder at LL accuracy in QCD can be described in 
terms of single-valued iterated integrals on the moduli space $\mathfrak{M}_{0,4}$ 
of Riemann spheres with four marked points, which are also given by 
SVHPLs~\cite{DelDuca:2013lma}. In this case, the single-valuedness 
can be traced back to the fact that neither of the momenta at the ends 
of the gluon ladder can vanish. That analysis can be extended to NLL 
accuracy for the BFKL ladder in QCD, as well as in full ${\cal N}=4$ 
SYM~\cite{DelDuca:2017peo}, however it requires a generalisation of 
the SVHPLs, recently introduced by Schnetz~\cite{Schnetz:2016fhy}. 
The control of the analytic structure of the BFKL ladder at NLL accuracy, 
and the freedom in defining its matter content, allowed Ref.~\cite{DelDuca:2017peo} 
to prove that there is no gauge theory of uniform and maximal transcendental 
weight which can match the maximal weight part of QCD, and to identify a 
set of conditions which allow to constrain the field content of theories 
for which the BFKL ladder has maximal weight.

\section{Regge phenomenology: three highlights}
\label{PhenRegge}

As is the case for theoretical developments, the field of phenomenological 
applications of Regge theory is too vast to be even superficially overviewed.
We will therefore confine ourselves to three examples, highlighting the 
fact that strong interactions display Regge behaviour both in their bound 
state structure and in the asymptotic behaviour of their cross sections; 
furthermore, high-energy logarithms may affect in a measurable way 
important multi-jet cross sections at LHC.

\begin{figure}
  \centerline{~~ \includegraphics[width=0.43\columnwidth]{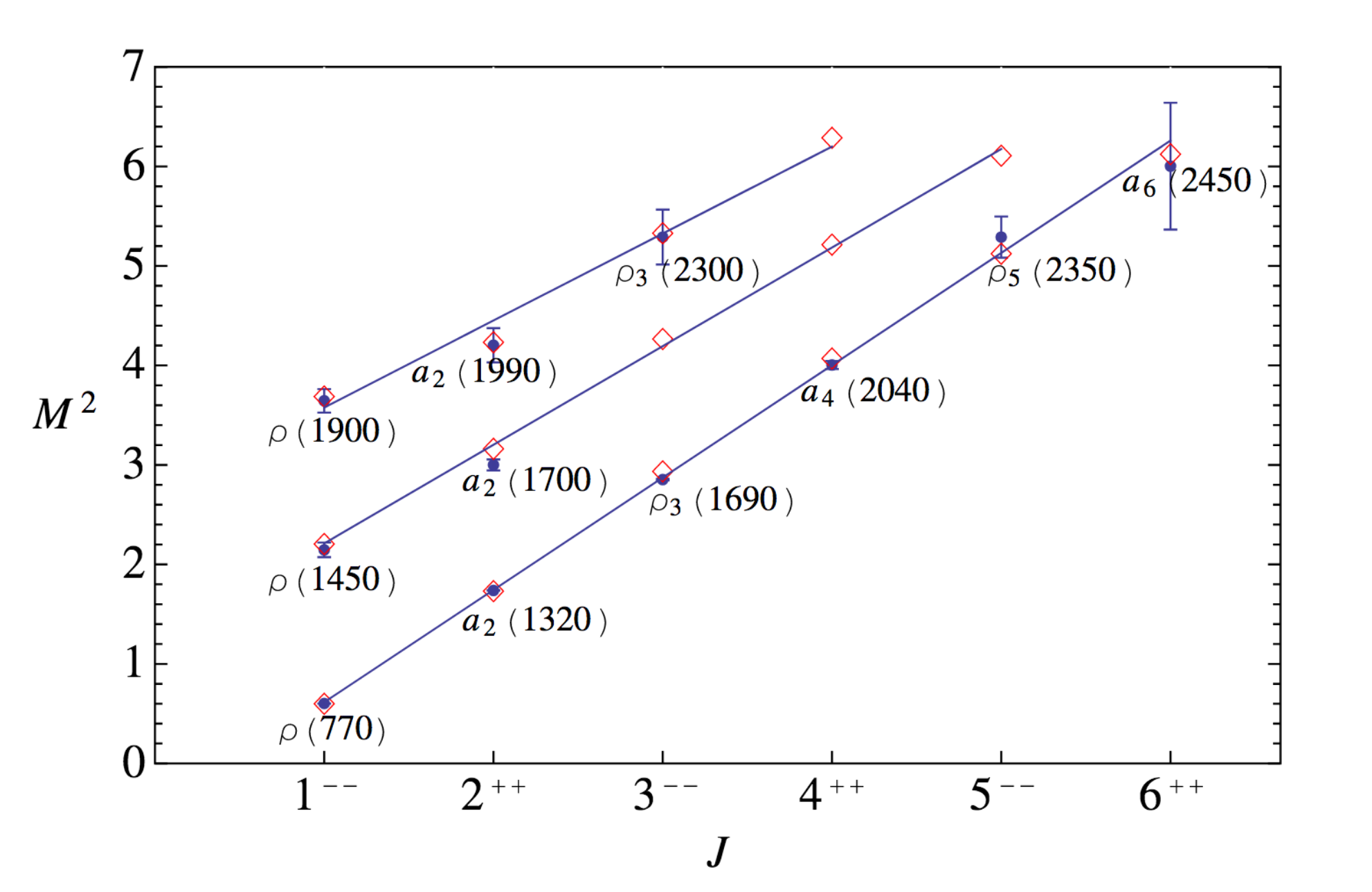} 
  ~~~~~~ \includegraphics[width=0.48\columnwidth]{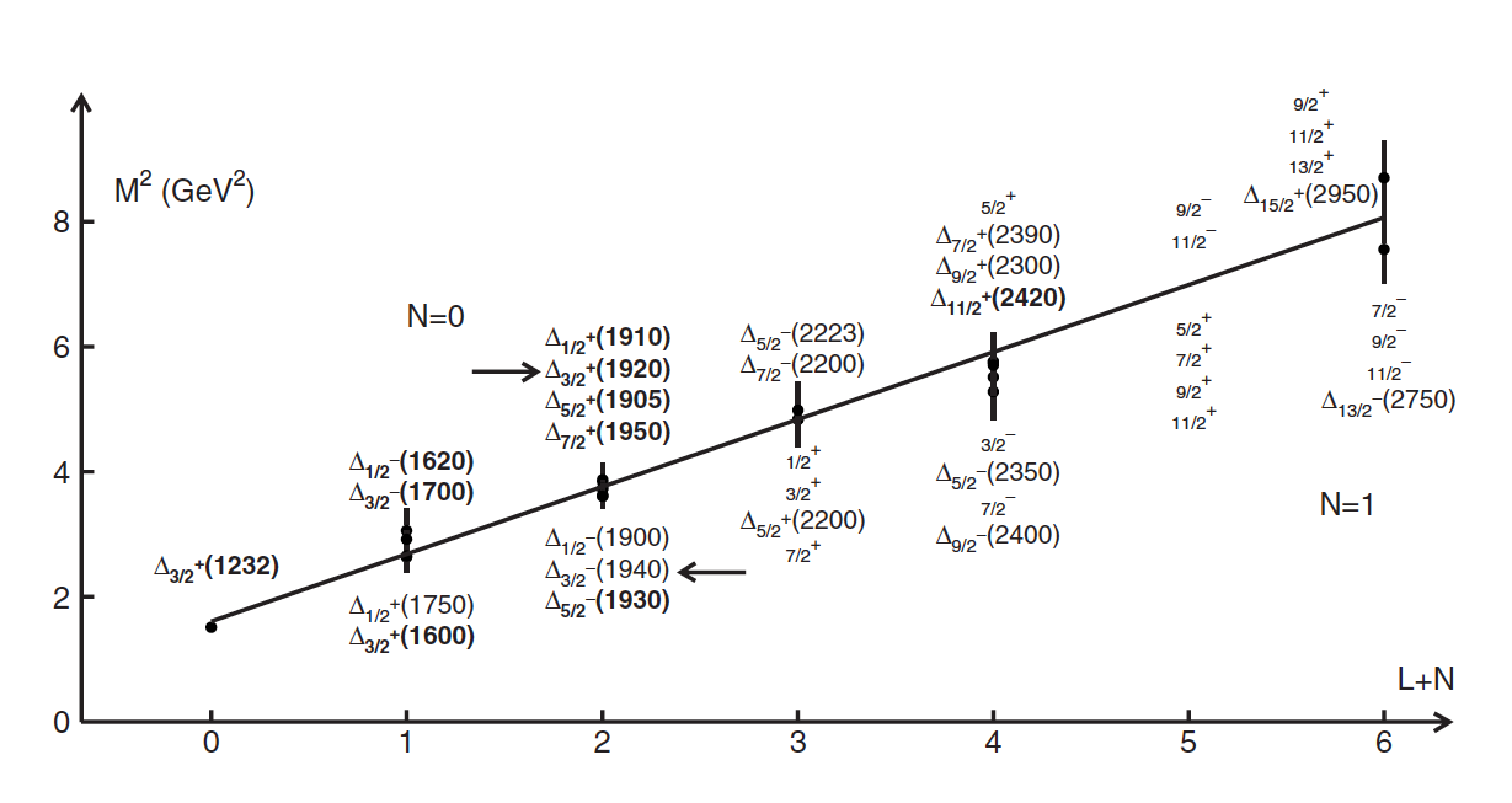} }
  \caption{In the first panel, a sample fit to the light isovector meson spectrum, 
  for the case of $\rho$ resonances, interpolated with linear Regge trajectories,
  from Ref.~\cite{Ebert:2009ub}; in the second panel, from Ref.~\cite{Klempt:2009pi}, 
  data for the $\Delta^*$ baryonic resonances, fit with a linear Regge trajectory 
  with parameters derived in Ref.~\cite{Forkel:2007cm}. } 
\label{FigMesBar}
\end{figure}

As a first example, in Fig.~\ref{FigMesBar}, we show two sample fits of hadron 
spectra, clearly displaying the alignment of meson and baryon states along 
linearly rising Regge trajectories. The first panel of Fig.~\ref{FigMesBar}, from
Ref.~\cite{Ebert:2009ub}, shows a set of isovector meson resonances (starting
with $\rho$ mesons), with the red diamonds representing predictions derived 
from a relativistic quark model. The second panel shows a baryon example,
taken from the review~\cite{Klempt:2009pi}, displaying resonances of the
$\Delta^*$ baryon; in this case the linearly rising Regge trajectory is not fit
from data, but it is derived from a non-perturbative model based on the
AdS/QCD correspondence~\cite{Forkel:2007cm}.
The existence of linearly rising trajectories was an early inspiration for the 
development of string theory, and effective string theories for a non-perturbative
description of strong interactions continue to be an active field of research.

\begin{figure}
  \centerline{~~~~~~~ \includegraphics[width=0.4\columnwidth]{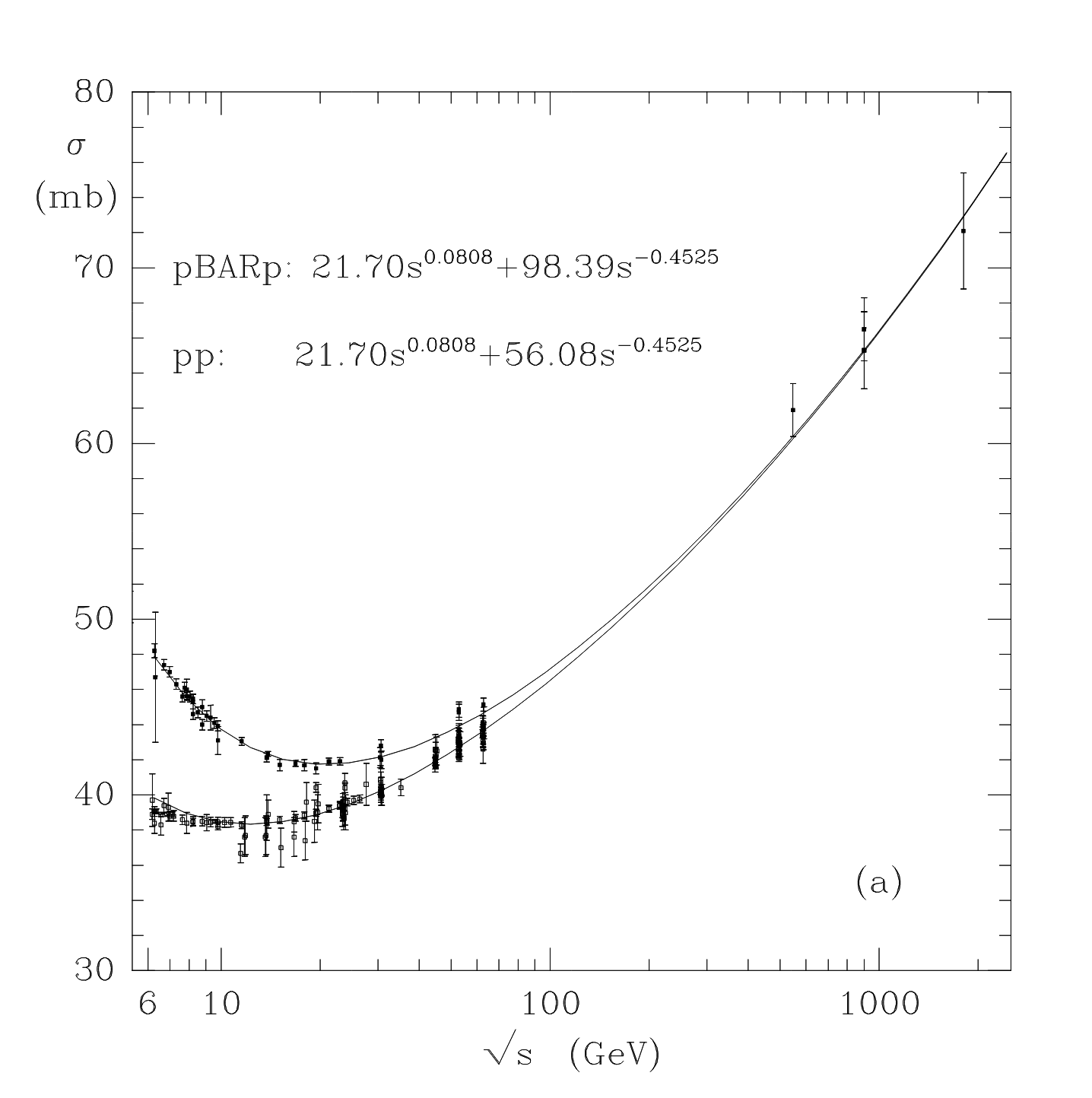} 
  ~~~~~ \includegraphics[width=0.5\columnwidth]{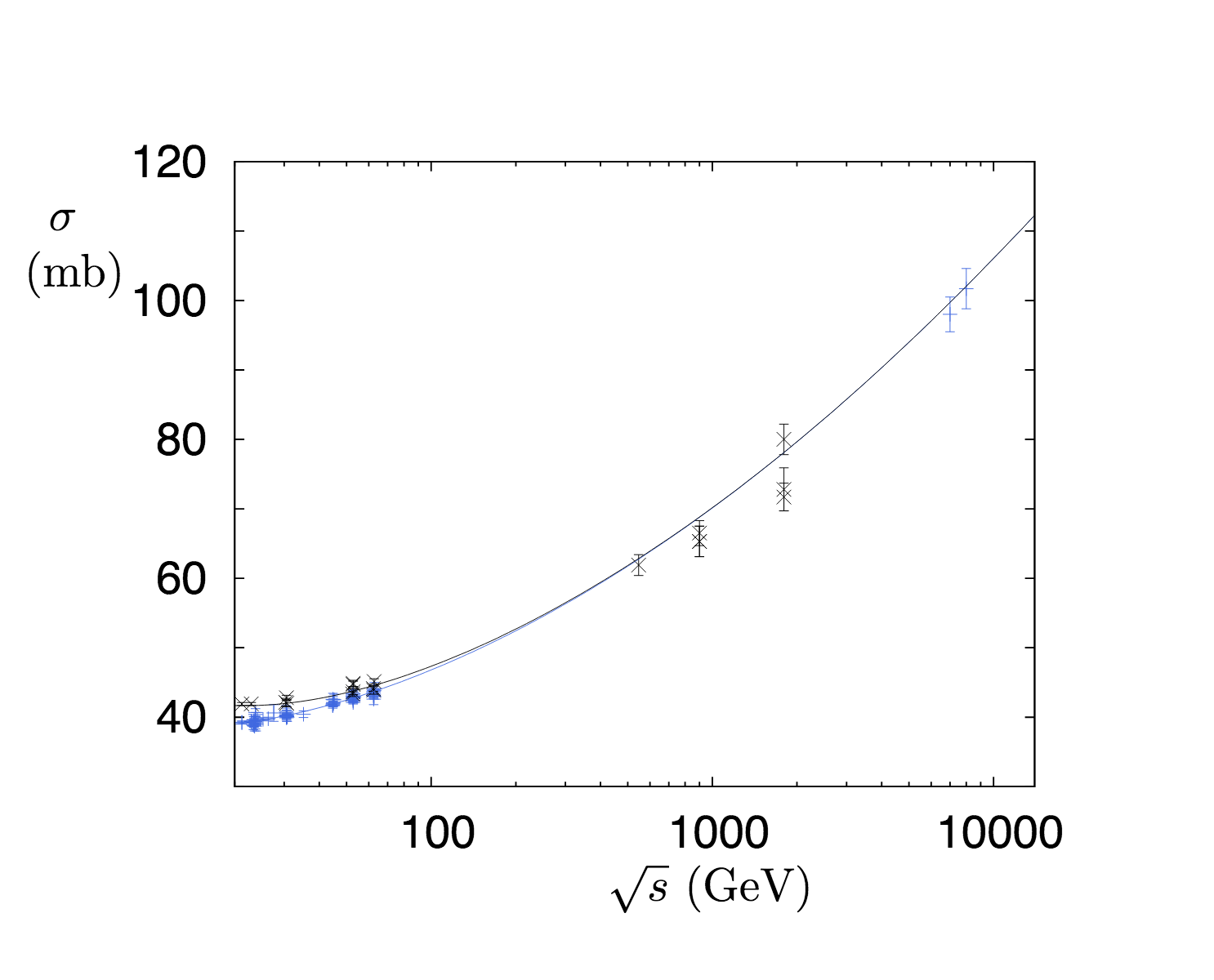} }
  \caption{In the first panel, the original Donnachie-Landshoff fit of the total 
  proton-proton and proton-antiproton cross sections with simple Regge 
  parametrisations, from  Ref.~\cite{Donnachie:1992ny}; in the second panel, 
  the 2013 update for the proton-proton case,  including early LHC data,
  from Ref.~\cite{Donnachie:2013xia}. } 
\label{figDL}
\end{figure}

As a second example, we display in Fig.~\ref{figDL} the well-known fits of total 
cross sections for proton-proton and proton-antiproton collisions by Donnachie and 
Landshoff. To quote the short abstract of their 1992 paper~\cite{Donnachie:1992ny},
the fits indeed show that ``Regge theory provides a very simple and economical
description of all total cross sections''. Indeed, Regge theory suggests that total 
hadronic cross section should be described at high energies by the exchange
of Reggeons, corresponding to sets of particles belonging to Regge trajectories.
Donnachie and Landshoff showed in Ref.~\cite{Donnachie:1992ny} that the
total cross sections data then available could be fit accordingly, using
simple expressions of the form
\beq
  \sigma^{\rm tot}_{ab} \, = \, X_{ab} \, s^\alpha \, + \, 
  Y_{ab} \, s^\beta \, ,
\label{DonnLand}
\eeq
where $a$ and $b$ denote the selected hadrons, and $X$ and $Y$ the 
corresponding fit parameters, while $\alpha \simeq 0.08$ and $\beta 
\simeq - 0.45$ are effective powers, roughly representing the contributions 
of the pomeron and $\rho$ trajectories: remarkably, the same $\alpha$ and 
$\beta$ can be used to describe many different processes, highlighting
the fact that the success of the fit depends on properties of the particles 
being exchanged in the scattering, rather than the particles being scattered.
The second panel of Fig.~\ref{figDL} shows an update of the fit, from
Ref.~\cite{Donnachie:2013xia}, which successfully includes the first LHC 
data.

\begin{figure}
  \centerline{\includegraphics[width=0.75\columnwidth]{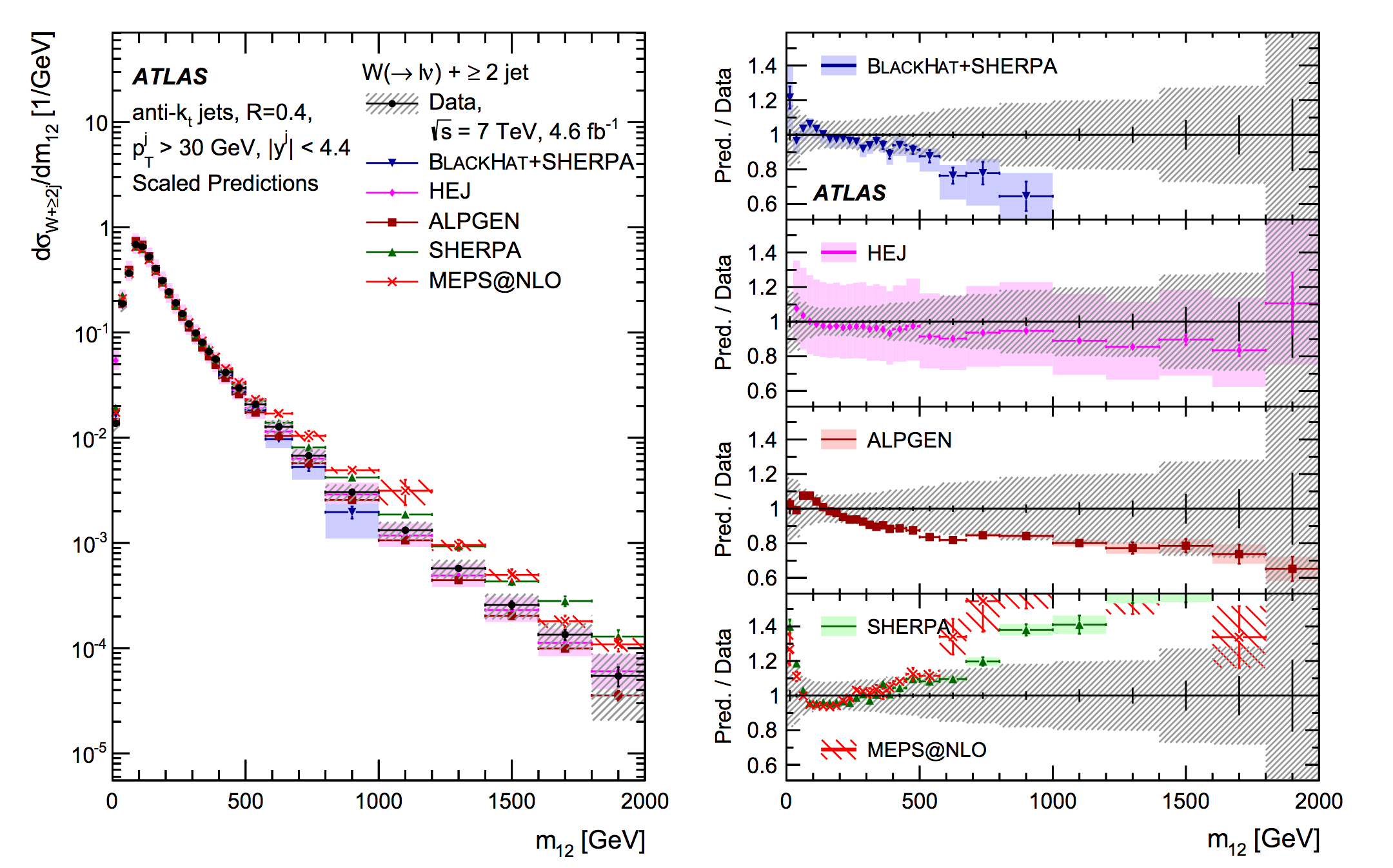} }
  \caption{ATLAS data giving the cross section for the production of $W$ 
  bosons in association with at least 2 jets, as a function of the di-jet invariant 
  mass of the two leading jets, compared with various theoretical predictions,
  from Ref.~\cite{Aad:2014qxa}. } 
\label{figHej}
\end{figure}

To conclude, we note that Regge theory provides, in principle, an all-order 
understanding of perturbative rapidity logarithms that arise in multi-particle
production cross sections, most notably in multi-jet cross sections currently
measured at LHC, when the jets are produced at large rapidity intervals.
Fig~\ref{figHej} shows a first successful attempt to fit such cross sections
including Regge contributions at leading logarithmic accuracy. In this case, 
the plot shows the cross section for the production of $W$ bosons in association 
with at least 2 jets, as a function of the di-jet invariant mass of the two leading 
jets. As shown in the second panel, the theoretical prediction using the 
High Energy Jet framework~\cite{Andersen:2011hs} successfully covers
the high invariant mass region of the data.
As future LHC measurements become more detailed and statistical errors are 
reduced as accumulated luminosity increases, it is very likely that the 
implementation of accurate high-energy resummations will become a 
necessary ingredient for data analysis. Many detailed theoretical results are 
already available in this direction (for a recent example, see~\cite{Bonvini:2018ixe}).

As we have tried to show with these brief notes, the legacy of Tullio Regge 
and his work on complex angular momenta will continue to play a central role 
in high-energy physics for a long time to come.

\subsubsection*{Acknowledgments}
We thank the Galileo Galilei Institute for Theoretical Physics in Firenze, and the Higgs Centre for Theoretical Physics in Edinburgh, for kind hospitality.

\end{document}